# Nonuniform Hydrodynamic Plasma-Wave Instability in a Gated Electron Channel

Yuhui Zhang†

Department of Electrical, Computer, and Systems Engineering, Rensselaer Polytechnic Institute, Troy, New York 12180, USA

† Corresponding-author. Current address: Sunnyvale, California, USA

**Abstract**

We report a theoretical and numerical study of plasma-wave instability in a current-driven InGaAs/GaAs gated electron channel. A finite drain current drives the channel away from a uniform plasma-wave cavity: the carrier density and plasma velocity decrease toward the drain, while the electron drift velocity increases. This current-induced spatial nonuniformity modifies both the propagation time and amplification of plasma waves and becomes increasingly important as the sonic regime is approached. Using a hydrodynamic description combined with the unified charge-control model, we determine the nonuniform steady state and analyze the resulting plasma-wave frequency and instability growth rate. The oscillation frequency is found to be governed by wave propagation through the entire channel rather than by the local drain conditions, leading to a pronounced current-induced frequency reduction that is not captured by a uniform-channel approximation. The instability results from the competition between boundary-induced plasma wave gain and distributed losses due to momentum relaxation and electron viscosity. The finite-viscosity theory shows good quantitative agreement with the simulation data over the subsonic regime and regularizes the singular behavior of the inviscid description near the sonic point. At higher currents, the dynamics become increasingly sensitive to the near-sonic structure and contact boundary conditions, indicating that the eventual suppression of the instability cannot be attributed to bulk hydrodynamics alone.

## 1. Introduction

The Dyakonov–Shur (DS) instability provides a mechanism for plasma-wave generation in a field-effect transistor driven by a DC drain current [1]. The same plasma-wave framework underlies resonant terahertz detection, frequency conversion, amplification, and current-driven emission in gated field-effect transistors [12–15]. The source and drain have different electrical boundary conditions. Consequently, the upstream and downstream waves can acquire a net amplitude gain after a round trip through the channel. Momentum relaxation and electron viscosity compete with this gain and modify the instability threshold [2,3,5,6]. The frequency and growth rate are therefore determined jointly by propagation through the electron channel and reflection at its terminals.

A current-driven transistor is generally not a uniform plasma-wave cavity. The DC electric field reduces the gate-to-channel voltage toward the drain. The carrier density decreases, whereas the drift velocity increases to maintain the imposed current. The plasma velocity also varies because it depends on the local charge compressibility. Spatial nonuniformity has been considered in earlier studies of instability and TeraFET detection [4,8]. Our previous analysis of current-driven TeraFETs used the UCCM to describe the DC state and identified the termination of the inviscid accelerating branch at the local sonic condition [7,11]. These results provide the starting point for the present analysis.

More broadly, the hydrodynamic description of electron transport and its viscous signatures have been developed theoretically and observed experimentally in several high-mobility two-dimensional systems [16–22]. These results motivate retaining the kinematic viscosity explicitly in the present plasma-wave model rather than representing all dissipation by a single local momentum-relaxation rate.

The oscillation frequency and growth rate require different levels of approximation. A phase integral can describe the frequency shift even when the channel is appreciably nonuniform. The growth rate additionally depends on the spatial evolution of the wave amplitudes. Replacing the channel by its drain values, or adding the local attenuation to an endpoint reflection factor without retaining the amplitude evolution, can therefore give misleading results near the sonic region. A numerical solution of the complete linearized problem is needed as a reference for the reduced expressions.

In this work, the derivation is kept in the physical variables $n_0(x)$, $v_0(x)$, $s(x)$, and $J_\mathrm{d}$. The Mach number is used only to describe the current bias and the sonic condition. Section 2 introduces the model and parameters. Section 3 derives the steady state. Section 4 develops the nonuniform frequency expressions from the local dispersion through the main reduced forms. Section 5 derives the finite-viscosity dispersion condition and the approximate growth-rate expressions. Section 6 discusses the sonic transition and the boundary limitation. Conclusions are given in Section 7. Additional detailed derivation can be found in the appendices.

## 2. Hydrodynamic Model

The one-dimensional model consists of the continuity equation and the Navier–Stokes equation for the gated two-dimensional electron gas,

$$\frac{\partial n}{\partial t}+\frac{\partial (nv)}{\partial x}=0, \tag{1}$$

$$\frac{\partial v}{\partial t}+v\frac{\partial v}{\partial x}+\gamma v+\frac{e}{m^*}\frac{\partial U}{\partial x}-\nu\frac{\partial^2 v}{\partial x^2}=0. \tag{2}$$

Here $n$ is the sheet carrier density, $v$ is the drift velocity, $m^*$ is the effective mass, and $e$ is the positive elementary charge. The momentum-relaxation rate is $\gamma = 1/\tau_{\mathrm{m}} = e/(\mu m^*)$, where $\mu$ is the mobility. The kinematic viscosity $\nu$ is constant within each calculation. The gate-to-channel voltage is $U = V_{\mathrm{gs}} - V_{\mathrm{th}} - V_{\mathrm{ch}}$. The temperature is fixed, and neither an electron energy equation nor a separate Poisson equation is included in the active model.

The UCCM relates the carrier density to the local voltage [7,9],

$$n(U)=\frac{C_{\mathrm{g}}\eta V_{\mathrm{T}}}{e}\ln\left[1+\exp\left(\frac{U}{\eta V_{\mathrm{T}}}\right)\right],\qquad V_{\mathrm{T}}=\frac{k_{\mathrm{B}}T}{e}. \tag{3}$$

Here $C_{\mathrm{g}}$ is the gate capacitance per unit area and $\eta$ is the ideality factor. Differentiating Eq. (3) gives the local plasma velocity directly,

$$s^2(n)=\frac{en}{m^*}\frac{\mathrm{d}U}{\mathrm{d}n}=\frac{e^2 n}{m^* C_{\mathrm{g}}}\left[1-\exp\left(-\frac{en}{C_{\mathrm{g}}\eta V_{\mathrm{T}}}\right)\right]^{-1}. \tag{4}$$

The strong-inversion gradual-channel approximation (GCA) result $s^2 = e^2 n/(m^* C_{\mathrm{g}})$ is recovered when $en/(C_{\mathrm{g}}\eta V_{\mathrm{T}}) \gg 1$. The full expression is used in the quantitative calculations; the GCA charge control is used only where stated to obtain closed analytical limits.

For current-driven operation without an externally applied AC signal, the electrical boundary conditions are

$$n(0,t)=N_{\mathrm{s}},\qquad e\,n(L,t)v(L,t)=J_{\mathrm{d}}. \tag{5}$$

The current $J_{\mathrm{d}}$ is the electrical current per unit width.

The above equations and boundary conditions are discretized and numerically solved in COMSOL Multiphysics® [10] using a 1D non-uniform mesh geometry. The COMSOL velocity interface also retains a zero diffusive flux at both outer endpoints. Since the coefficient of diffusion is $\nu$ and the convection term is entered separately, its smooth continuum interpretation is

$$\left.\frac{\partial v}{\partial x}\right|_{\mathrm{x=0}} = 0, \qquad \left.\frac{\partial v}{\partial x}\right|_{\mathrm{x=L}} = 0. \tag{6}$$

These saved settings are reported separately from the analytical contact closure. In the finite-viscosity bulk calculation below, only the source velocity-gradient condition is used with Eq. (5); the compatibility of the additional drain gradient condition is discussed in Section 6. The residual weak contribution used for numerical stabilization is not included in the continuum theory. It is proportional to the continuity residual and vanishes for an exact solution of Eq. (1).

The time-dependent simulation starts from uniform fields,

$$n(x,0) = N_{\mathrm{s}}, \qquad v(x,0) = K_{\mathrm{v}}S, \qquad J_{\mathrm{d}} = eN_{\mathrm{s}}K_{\mathrm{v}}S. \tag{7}$$

The source AC term is multiplied by zero in the saved model. The initial field should therefore be distinguished from the nonuniform DC background used for eigenvalue analysis. Since momentum relaxation is nonzero, the uniform initial state must first adjust to the current-driven channel. An exponential fit to an early transient is not automatically a measurement of a single linear eigenmode.

Table I. Parameters of the current InGaAs/GaAs model.

| **Parameter** | **Value** | **Parameter** | **Value** |
| --- | --- | --- | --- |
| $m^*/m_{\mathrm{e}}$ | 0.041 | $L$ | 130 nm |
| $V_{\mathrm{g0}},\ V_{\mathrm{th}}$ | 1.6 V, 1.0 V | $T,\ \eta$ | 300 K, 3 |
| $C_{\mathrm{g}}$ | $5.36 \times 10^{-3}\ \mathrm{F/m^2}$ | $N_{\mathrm{s}}$ | $2 \times 10^{16}\ \mathrm{m^{-2}}$ |
| $\mu$ | $1.2\ \mathrm{m^2/(V\,s)}$ | $\gamma$ | $3.57 \times 10^{12}\ \mathrm{s^{-1}}$ |
| $S$ | $1.6 \times 10^{6}\ \mathrm{m/s}$ | $\tau_{\mathrm{m}}$ | 0.288 ps |
| $\nu$ | $0.8,\ 2,\ 20,\ 100\ \mathrm{cm^2/s}$ | $M_{\mathrm{s}}$ | $v_0(0)/s(0)$ |

The imposed source density is $N_{\mathrm{s}} = C_{\mathrm{g}}\big(V_{\mathrm{g0}} - V_{\mathrm{th}}\big)/e$, whereas the saved velocity scale $S$ is evaluated from the full UCCM at the source voltage. Evaluating Eq. (4) at the imposed density gives $s(0) = 1.604683 \times 10^{6}\ \mathrm{m/s}$, only about $0.0028\%$ below $S$. The retained numerical curves use the saved scale and identify $M_{\mathrm{s}} \simeq K_{\mathrm{v}}$ at this accuracy. The source charge ratio is $eN_\mathrm{s}/(C_\mathrm{g}\eta V_\mathrm{T}) = 7.73635$. The present revision retains these parameters and the numerical curves of the parameter-updated calculation; the earlier nominal values are not restored.

**3. Current-driven nonuniform steady state**

We first obtain the DC state before considering its oscillations. In the steady state, Eq. (1) gives

$$J_\mathrm{d} = e\,n_0(x)v_0(x), \qquad n_0(x) = \frac{J_\mathrm{d}}{e\,v_0(x)}. \tag{8}$$

The density and velocity gradients are consequently related by

$$\frac{\mathrm{d}n_0}{\mathrm{d}x} = -\frac{n_0}{v_0}\frac{\mathrm{d}v_0}{\mathrm{d}x}, \qquad \frac{e}{m^*}\frac{\mathrm{d}U_0}{\mathrm{d}x} = -\frac{s^2}{v_0}\frac{\mathrm{d}v_0}{\mathrm{d}x}. \tag{9}$$

Substitution into the steady momentum equation gives

$$\left(1-\frac{s^2}{v_0^2}\right)\frac{\mathrm{d}v_0}{\mathrm{d}x} - \frac{\nu}{v_0}\frac{\mathrm{d}^2 v_0}{\mathrm{d}x^2} + \gamma = 0. \tag{10}$$

This equation separates the three physical contributions: current acceleration, viscous momentum diffusion, and momentum relaxation. The plasma velocity is not an independently prescribed profile. Combining Eqs. (4) and (8) gives its explicit dependence on the local drift velocity,

$$s^2(v_0) = \frac{eJ_\mathrm{d}}{m^* C_\mathrm{g} v_0}\left[1-\exp\left(-\frac{J_\mathrm{d}}{C_\mathrm{g}\eta V_\mathrm{T} v_0}\right)\right]^{-1}. \tag{11}$$

For the source-compatible finite-viscosity branch, Eq. (10) is solved with

$$v_0(0) = v_\mathrm{s} = \frac{J_\mathrm{d}}{eN_\mathrm{s}}, \qquad v_0'(0) = 0. \tag{12}$$

A prime on a spatial field denotes differentiation with respect to $x$. On the accelerating subsonic branch, $v_0$ rises while $n_0$ and $s$ decrease toward the drain. The local ratio $M(x) = v_0(x)/s(x)$ therefore increases more rapidly than the velocity alone. Fig. 1 illustrates this current-induced nonuniformity on the inviscid branch.

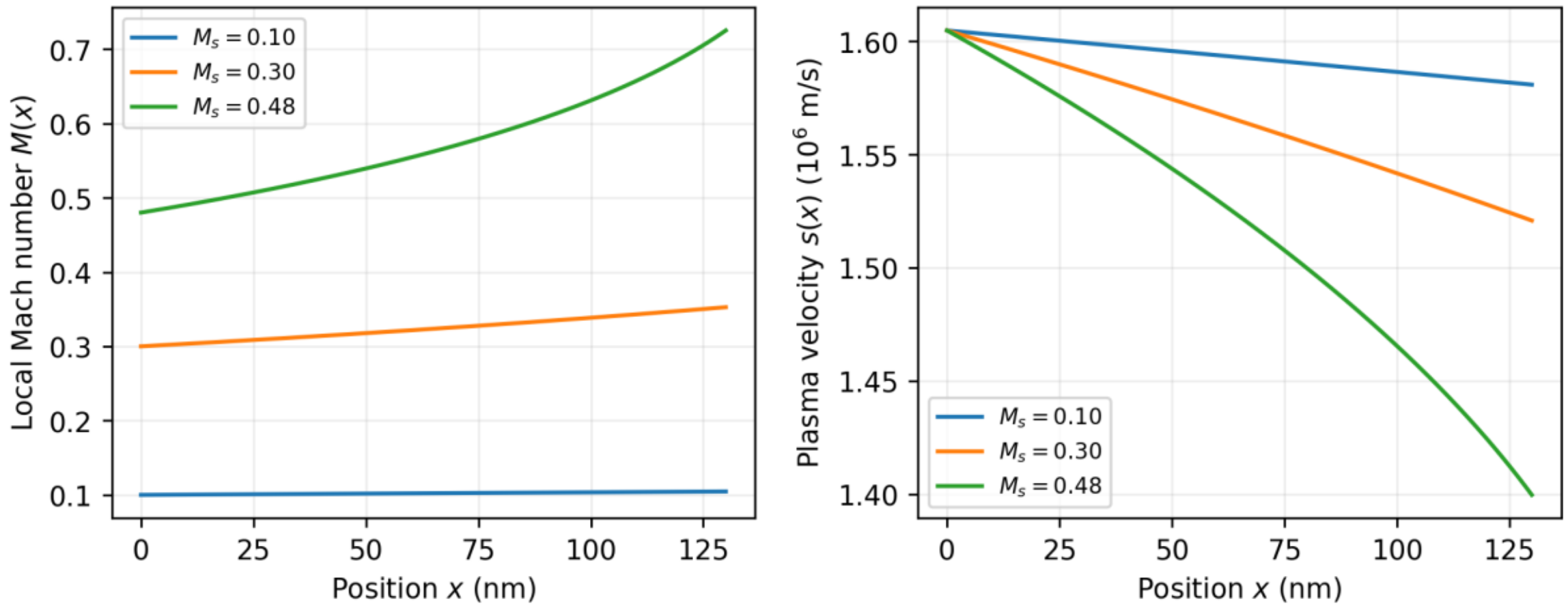

Fig. 1. DC profiles calculated from the inviscid UCCM steady state. The current accelerates the carriers toward the drain while reducing the plasma velocity. Spatial nonuniformity becomes pronounced as the source Mach number approaches the sonic regime.

When $\nu = 0$, Eq. (10) is first order and can be integrated in the physical velocity,

$$x = \frac{1}{\gamma} \int_{\mathrm{v_s}}^{v_0(x)} \left[ \frac{s^2(v)}{v^2} - 1 \right] \mathrm{d}v. \tag{13}$$

The regular accelerating branch reaches its endpoint when $v_\mathrm{d} = s_\mathrm{d}$, where $v_\mathrm{d} = v_0(L)$ and $s_\mathrm{d} = s(L)$. Solving Eq. (13) together with this condition gives $M_{\mathrm{s,sonic}} \simeq 0.51198$ for the retained parameter convention. This is a steady-state sonic condition, not the condition $\omega_2 = 0$ for a dynamical instability.

### 3.1 Strong-inversion limit and the nonuniformity factor

The steady relation becomes particularly transparent under strong inversion. Defining the physical choking velocity $v_\mathrm{cr}$ gives

$$s^2 v_0 = v_\mathrm{cr}^3, \qquad v_\mathrm{cr}^3 = \frac{eJ_\mathrm{d}}{m^* C_\mathrm{g}}. \tag{14}$$

Equation (13) then yields the convection-inclusive current relation [7,11],

$$\frac{v_\mathrm{cr}^3}{2v_0^2(x)} + v_0(x) = \frac{v_\mathrm{cr}^3}{2v_\mathrm{s}^2} + v_\mathrm{s} - \gamma x. \tag{15}$$

The commonly used Mach representation follows directly, without introducing normalized density or channel coordinates,

$$v_0 = v_{\mathrm{cr}} M^{2/3}, \qquad s = v_{\mathrm{cr}} M^{-1/3},$$
$$\frac{\mathrm{d}x}{\mathrm{d}M} = \frac{2v_{\mathrm{cr}}}{3\gamma}\frac{1-M^2}{M^{7/3}}. \tag{16}$$

The factor $1 - M^2$ will cancel the large local phase density in the frequency integral. This cancellation is the reason that a narrow near-sonic region does not force the total cavity frequency to vanish. Equations (14)–(16) are closed-form limits; the numerical profiles and updated comparison use Eqs. (10) and (11) with the complete UCCM.

## 4. Oscillation frequency in a nonuniform channel

### 4.1 Local viscous dispersion and the cubic correction

Locally freeze the DC coefficients and write the perturbation as $\exp(ikx - i\omega t)$. The continuity and momentum equations become

$$(\omega - kv_0)n_1 = kn_0 v_1,$$
$$\left[\omega - kv_0 + i(\gamma + \nu k^2)\right]v_1 = k\frac{s^2}{n_0}n_1. \tag{17}$$

Eliminating either perturbation gives the local dispersion relation

$$\left[\omega - kv_0 + i(\gamma + \nu k^2)\right](\omega - kv_0) = s^2k^2. \tag{18}$$

Expansion in $k$ makes the effect of finite drift and viscosity explicit,

$$-i\nu v_0 k^3 + (v_0^2 - s^2 + i\nu\omega)k^2$$
$$-v_0(2\omega + i\gamma)k + \omega(\omega + i\gamma) = 0. \tag{19}$$

The term proportional to $\nu v_0 k^3$ adds a third spatial root. To obtain a tractable two-wave approximation, first omit only this cubic term and define the remaining polynomial

$$Q(k) = (v_0^2 - s^2 + i\nu\omega)k^2 - v_0(2\omega + i\gamma)k + \omega(\omega + i\gamma). \tag{20}$$

The two quadratic roots are

$$k_{\pm}^{(Q)} = \frac{v_0(\omega + i\gamma/2) \pm \sqrt{(\omega^2 + i\gamma\omega)(s^2 - i\nu\omega) - v_0^2\gamma^2/4}}{v_0^2 - s^2 + i\nu\omega}. \tag{21}$$

The signs in Eq. (21) label the algebraic branches, not the direction of propagation. With a positive real frequency and $0 < v_0 < s$, the chosen square root approaches $\omega s$ as damping vanishes. Hence $k_+^{(Q)} \to -\omega/(s - v_0)$ is the upstream root and $k_-^{(Q)} \to \omega/(s + v_0)$ is the downstream root. This convention is used consistently in the phase and gain formulas below.

Writing the complete dispersion as $Q(k) - i\nu v_0 k^3 = 0$ and expanding about a quadratic root gives

$$Q'\left(k_\pm^{(Q)}\right)\delta k_\pm - i\nu v_0\left(k_\pm^{(Q)}\right)^3 \simeq 0, \tag{22}$$

$$\delta k_\pm = \frac{i\nu v_0\left(k_\pm^{(Q)}\right)^3}{2(v_0^2 - s^2 + i\nu\omega)k_\pm^{(Q)} - v_0(2\omega + i\gamma)}, \tag{23}$$
$$\tilde{k}_\pm = k_\pm^{(Q)} + \delta k_\pm.$$

The derivative of $Q$ in Eq. (22) is with respect to $k$. This correction is first order in the omitted cubic term; it is not a uniformly valid expansion in $\nu$ at every drift velocity. Its local validity requires

$$\max_{0\le x\le L}\left|\frac{\delta k_\pm}{k_\pm^{(Q)}}\right| \ll 1. \tag{24}$$

The upstream root generally gives the more restrictive test near the sonic region. The exact finite-viscosity eigenvalue problem in Section 5 does not eliminate this third degree of freedom and does not require Eq. (24).

### 4.2 Global phase condition

The cavity frequency follows from propagation through the entire nonuniform channel. In the leading two-wave approximation, the density perturbation is a sum of the two phase factors. The source condition makes their source amplitudes opposite. From Eq. (17), the perturbation of particle current on each local branch is $\omega n_1/k$. Applying the fixed-current drain condition therefore gives

$$\exp\left[i\int_0^L (k_+ - k_-)\ \mathrm{d}x\right] = \frac{k_+(L,\omega)}{k_-(L,\omega)}. \tag{25}$$

This step neglects the separate spatial transport of the wave amplitudes. It is a leading WKB construction, not an exact boundary condition for arbitrary spatial gradients. Taking a logarithm gives the common complex phase–amplitude equation

$$G(\omega) = \ln\left[\frac{k_+(L,\omega)}{k_-(L,\omega)}\right] - i\int_0^L (k_+ - k_-)\ \mathrm{d}x - 2\pi i\mathrm{p} = 0. \tag{26}$$

The integer p selects a continuous logarithm branch. The real-frequency phase condition is

$$-\mathrm{Re}\int_0^L [k_+(x,\omega_1) - k_-(x,\omega_1)]\ \mathrm{d}x = \Phi(\omega_1),$$
$$\Phi = 2\pi \mathrm{p} - \arg\left[\frac{k_+(L,\omega_1)}{k_-(L,\omega_1)}\right] \simeq q\pi, \qquad q = 1,3,5,\ldots \tag{27}$$

The fundamental mode has $q = 1$. In the comparison curves the endpoint phase is approximated by $q\pi$, as in the earlier frequency hierarchy. If it is retained, the same endpoint expression must be used throughout the calculation. The general quadratic form of Eq. (27) is

$$\mathrm{Re}\int_0^L \frac{2\sqrt{(\omega_1^2 + i\gamma\omega_1)(s^2 - i\nu\omega_1) - v_0^2\gamma^2/4}}{s^2 - v_0^2 - i\nu\omega_1}\ \mathrm{d}x = \Phi(\omega_1). \tag{28}$$

The highest two-root frequency approximation restores Eq. (23) before accumulating the phase,

$$-\mathrm{Re}\int_0^L \left[k_+^{(Q)} - k_-^{(Q)} + \delta k_+ - \delta k_-\right]_{\omega=\omega_1} \mathrm{d}x = \Phi(\omega_1). \tag{29}$$

Equations (28) and (29) retain the nonuniform coefficients throughout the channel. They are implicit expressions for $\omega_1$; the real trial frequency is adjusted until the total phase condition is satisfied. The full quadratic expression retains the mixed friction–viscosity and complex phase terms that are omitted in the reductions below.

### 4.3 Principal reduced forms of the oscillation frequency

When $|\nu\omega_1| \ll s^2 - v_0^2$ and the imaginary part of the discriminant is a perturbation, Eq. (28) reduces to

$$2\int_0^L \frac{\sqrt{\omega_1^2(s^2 + \gamma\nu) - v_0^2\gamma^2/4}}{s^2 - v_0^2}\ \mathrm{d}x \simeq q\pi. \tag{30}$$

This is the finite-$\gamma^2$ reduction, called Level III in the numerical comparison. It is not the complete second-order expansion in friction: other complex-discriminant terms have been discarded. Neglecting the explicit $v_0^2\gamma^2/4$ term gives the Level II expression,

$$\omega_1^{(\mathrm{II})} \simeq \frac{q\pi}{2\int_0^L \frac{\sqrt{s^2(x) + \gamma\nu}}{s^2(x) - v_0^2(x)}\ \mathrm{d}x}. \tag{31}$$

If the direct $\gamma\nu$ correction is also neglected in the AC phase, the result becomes

$$\omega_1^{(\mathrm{I})} \simeq \frac{q\pi}{2\int_0^L \frac{s(x)}{s^2(x) - v_0^2(x)}\ \mathrm{d}x}. \tag{32}$$

The friction-induced nonuniform DC profile remains in Eq. (32); only the direct dissipative corrections to the local AC phase have been dropped. The denominator is the round-trip propagation time,

$$\mathrm{T} = \int_0^L \left[\frac{1}{s+v_0} + \frac{1}{s-v_0}\right] \mathrm{d}x = 2\int_0^L \frac{s}{s^2 - v_0^2}\,\mathrm{d}x. \tag{33}$$

Thus Eq. (32) is simply $\omega_1^{(\mathrm{I})} = q\pi/\mathrm{T}$. It makes the physics explicit: the frequency measures a global propagation time, not the value of $s^2 - v_0^2$ at a single drain point. For the source-matched uniform-channel reference used in Fig. 2, the uniform plasma velocity and drift are fixed at their source values, giving

$$\omega_1^{\mathrm{uniform}} \simeq \frac{q\pi s_{\mathrm{s}}\left(1 - {M_{\mathrm{s}}}^2\right)}{2L}. \tag{34}$$

Using Eq. (16), the integrand on the inviscid strong-inversion background satisfies

$$\frac{s}{s^2 - v_0^2}\frac{\mathrm{d}x}{\mathrm{d}M} = \frac{2}{3\gamma M^2}. \tag{35}$$

The factors $1 - M^2$ cancel exactly. Integration therefore gives the explicit nonuniform expression

$$\omega_1^{(\mathrm{I})} \simeq \frac{3q\pi\gamma}{4\left(M_{\mathrm{s}}^{-1} - M_{\mathrm{d}}^{-1}\right)}. \tag{36}$$

Unlike a drain-local substitution in Eq. (34), Eq. (36) remains finite as $M_{\mathrm{d}} \to 1$ at fixed nonzero friction. Retaining the direct $\gamma\nu$ contribution gives the corresponding strong-inversion closed form

$$\omega_1^{(\mathrm{II})} \simeq \frac{3q\pi\gamma}{4\left[\frac{(1+\gamma\nu/s_{\mathrm{s}}^2)^{3/2}}{M_{\mathrm{s}}} - \frac{(1+\gamma\nu/s_{\mathrm{d}}^2)^{3/2}}{M_{\mathrm{d}}}\right]}. \tag{37}$$

Here $s_{\mathrm{s}} = s(0)$ and $s_{\mathrm{d}} = s(L)$. The integration leading to Eq. (37) is given in Appendix B without introducing a normalized velocity or an auxiliary charge variable. Equations (36) and (37) are the strong-inversion limits of Eqs. (32) and (31); they are not substituted for the full UCCM in the updated numerical curves.

To expose the next correction, expand the square root in Eq. (30) about its Level II value. To first order in the displayed $\gamma^2$ term,

$$\omega_1^{(\mathrm{III})} \simeq \omega_1^{(\mathrm{II})} + \frac{\gamma^2}{4q\pi}\int_0^L \frac{v_0^2\,\mathrm{d}x}{(s^2 - v_0^2)\sqrt{s^2 + \gamma\nu}}. \tag{38}$$

The integral is positive on the regular subsonic branch, so this particular correction raises the frequency. When $\nu = 0$ and strong inversion is used, Eq. (38) further reduces to

$$\omega_1^{(\mathrm{III})} \simeq \frac{3q\pi\gamma}{4\left(M_\mathrm{s}^{-1} - M_\mathrm{d}^{-1}\right)} + \frac{\gamma(M_\mathrm{d} - M_\mathrm{s})}{6q\pi}. \tag{39}$$

The quadratic equation obtained before linearizing the frequency correction is retained in Appendix B. Finally, restoring all quadratic phase terms and the cubic root correction yields Eq. (29), referred to as Level IV. These levels isolate particular physical contributions; they should not be read as a uniform sequence of powers of $\nu$ near the sonic point.

### 4.4 Comparison of the frequency approximations

Fig. 2 compares the four nonuniform frequency levels, the source-matched uniform-channel reference of Eq. (34), and the simulation values. The exact-UCCM inviscid DC profile is used for this frequency hierarchy, while viscosity is retained in the AC roots according to the stated level. This is distinct from Section 5, where the background itself depends on viscosity. Table II reports the updated errors on the unchanged comparison set $M_\mathrm{s} \leq 0.48$. The source-matched uniform reference increasingly overestimates the frequency as the current increases, whereas the nonuniform phase-integral curves capture the pronounced current-induced frequency reduction and remain substantially closer to the simulation data. The largest improvement relative to a drain-local estimate is associated with the nonuniform phase integral. The cubic correction is most visible at the largest viscosity. No new parameter fitting is performed in restoring these derivations.

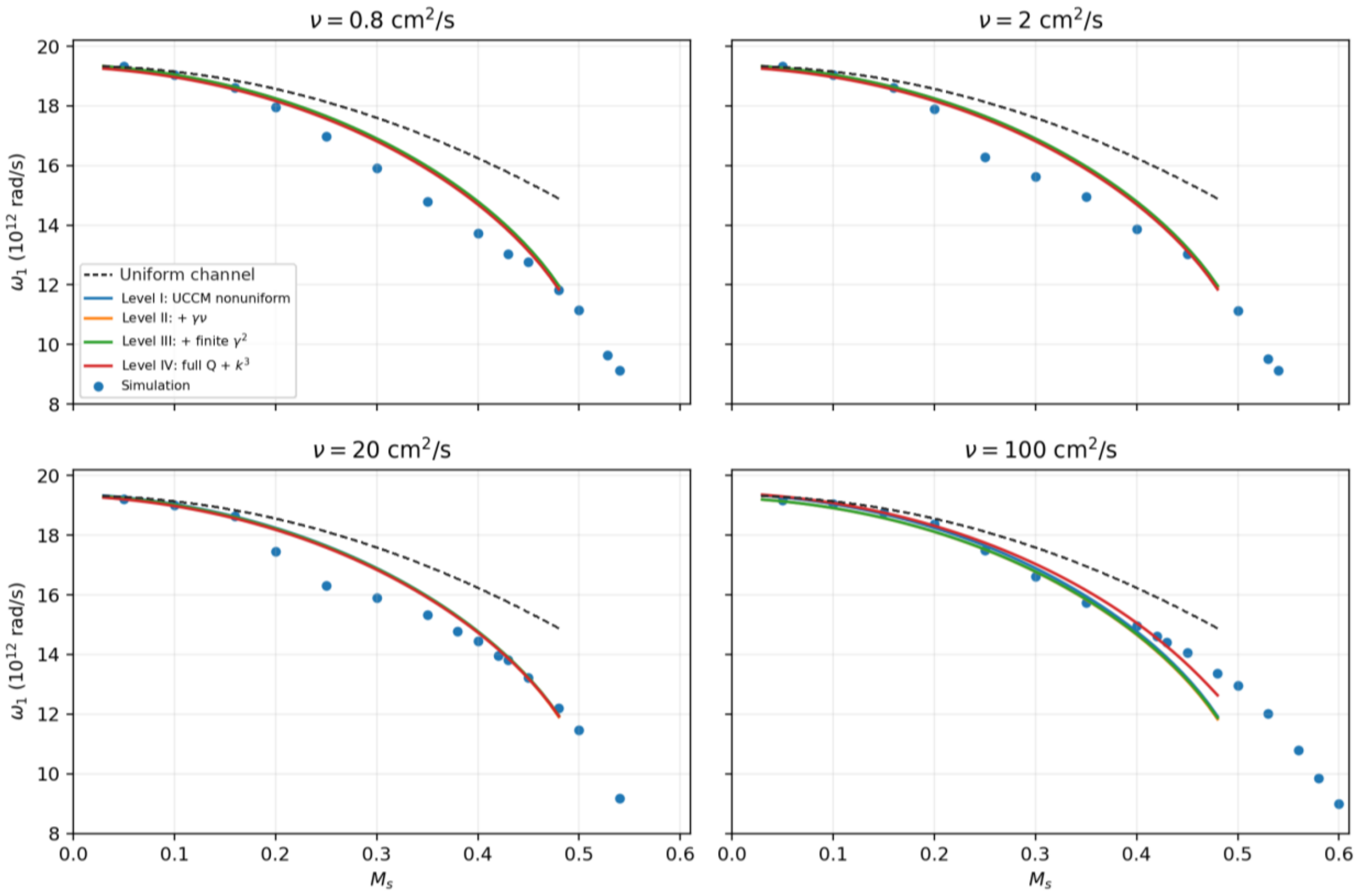

Fig. 2. Comparison of the nonuniform frequency hierarchy and the source-matched uniform-channel reference with the simulation data. Levels I, II, and III follow Eqs. (32), (31), and (30), respectively; Level IV follows Eq. (29). The dashed black curve is the uniform-channel result of Eq. (34). The updated UCCM background and model parameters are used throughout. Points above the analytical comparison range are shown for context.

Table II. Frequency RMSE for the available points with $M_s \leq 0.48$. All entries are in $10^{12}$ rad/s.

| $\nu$ (cm$^2$/s) | **Level I** | **Level II** | **Level III** | **Level IV** |
|---|---|---|---|---|
| 0.8 | 0.672 | 0.671 | 0.685 | 0.612 |
| 2 | 0.778 | 0.776 | 0.786 | 0.722 |
| 20 | 0.577 | 0.562 | 0.568 | 0.543 |
| 100 | 0.533 | 0.586 | 0.566 | 0.308 |

## 5. Instability growth rate in the nonuniform channel

### 5.1 Complete finite-viscosity eigenvalue problem

The phase condition alone does not determine the accuracy of the growth rate. We therefore first derive the full nonuniform small-signal problem, before reducing it to approximate gain–loss formulas. Write the perturbations as

$$n = n_0 + n_1(x)e^{-i\omega t}, \qquad v = v_0 + v_1(x)e^{-i\omega t}, \\ \omega = \omega_1 + i\omega_2. \tag{40}$$

The physical fields are the real parts of these expressions. A positive $\omega_2$ denotes temporal growth. Unlike the local derivation of Eq. (18), spatial derivatives of all DC coefficients are now retained. Linearizing Eqs. (1) and (2) gives

$$-i\omega n_1 + \frac{\mathrm{d}}{\mathrm{d}x}(v_0 n_1 + n_0 v_1) = 0, \tag{41}$$

$$-i\omega v_1 + v_0 v_1' + (v_0' + \gamma)v_1 - \nu v_1'' + \frac{\mathrm{d}}{\mathrm{d}x}\left(\frac{s^2}{n_0}n_1\right) = 0. \tag{42}$$

The last derivative in Eq. (42) includes both the derivative of the perturbation and that of the UCCM coefficient. It must not be replaced by $(s^2/n_0)n_1'$ in a nonuniform channel. The DC profiles used here are obtained from Eq. (10) at the same viscosity as the perturbation problem. The electrical boundary conditions and the source-compatible velocity condition linearize to

$$n_1(0) = 0, \qquad v_1'(0) = 0, \\ v_\mathrm{d} n_1(L) + n_\mathrm{d} v_1(L) = 0. \tag{43}$$

For numerical integration, Eq. (41) first gives

$$n_1' = \frac{i\omega - v_0'}{v_0} n_1 - \frac{n_0'}{v_0} v_1 - \frac{n_0}{v_0} v_1'. \tag{44}$$

Substituting this result into Eq. (42) gives the second derivative of the velocity perturbation explicitly,

$$\nu v_1'' = \left[\left(\frac{s^2}{n_0}\right)' + \frac{s^2(i\omega - v_0')}{n_0 v_0}\right] n_1 \\ + \left[v_0' + \gamma - i\omega - \frac{s^2 n_0'}{n_0 v_0}\right] v_1 + \left(v_0 - \frac{s^2}{v_0}\right) v_1'. \tag{45}$$

Equations (44) and (45) form three first-order spatial equations in $n_1$, $v_1$, and $v_1'$. They do not contain a denominator $s^2 - v_0^2$, so finite viscosity removes the explicit sonic denominator of the inviscid spatial system. The calculation still requires a regular DC state and a specified contact closure.

To obtain a scalar eigenvalue condition, integrate a source-normalized solution with $n_1(0) = 0$, $v_1(0) = v_{\mathrm{ref}}$, and $v_1'(0) = 0$, where the nonzero reference velocity is arbitrary. The remaining drain-current residual is

$$\mathrm{R}_{\mathrm{v}}(\omega) = e[v_{\mathrm{d}} n_1(L,\omega) + n_{\mathrm{d}} v_1(L,\omega)]. \tag{46}$$

The exact finite-viscosity eigenvalue condition within this boundary model is therefore

$$\mathrm{R}_{\mathrm{v}}(\omega_*) = 0, \qquad \omega_1 = \mathrm{Re}[\omega_*], \qquad \omega_2 = \mathrm{Im}[\omega_*]. \tag{47}$$

The scalar normalization of the source solution does not change the zeros of Eq. (47). The equation provides the nonuniform finite-viscosity expression used for the theory curves in Fig. 3. It retains the full linear amplitude evolution and does not assume small $|\omega_2|/\omega_1$ or eliminate a viscous root. It is an implicit spectral expression, rather than an elementary closed formula.

### 5.2 Inviscid transfer-matrix limit

The inviscid limit provides an independent reference. Define the electrical current perturbation and use the linearized continuity equation,

$$J_1 = e(n_0 v_1 + v_0 n_1), \qquad J_1' = ie\omega n_1. \tag{48}$$

Multiplying the inviscid momentum equation by $n$ before linearizing yields

$$\frac{\mathrm{d}}{\mathrm{d}x}\left[(s^2 - v_0^2)n_1 + \frac{2v_0 J_1}{e}\right] + \frac{\gamma - i\omega}{e} J_1 = 0. \tag{49}$$

Combining Eqs. (48) and (49) gives

$$n_1' = -\frac{(s^2 - v_0^2)' + 2i\omega v_0}{s^2 - v_0^2} n_1 + \frac{i\omega - \gamma - 2v_0'}{e(s^2 - v_0^2)} J_1, \qquad J_1' = ie\omega n_1. \tag{50}$$

All coefficients vary with position. In particular, $(s^2 - v_0^2)'$ is retained, so this propagation includes the nonuniform amplitude transport that is absent from the simple local-wave matching. If $\mathbf{T}$ denotes the resulting two-state transfer matrix,

$$\begin{pmatrix} n_1(L) \\ J_1(L) \end{pmatrix} = \mathbf{T}(L,0;\omega) \begin{pmatrix} n_1(0) \\ J_1(0) \end{pmatrix}. \tag{51}$$

The source-density and drain-current conditions then give

$$T_{22}(L,0;\omega_*) = 0, \qquad \omega_2 = \mathrm{Im}[\omega_*]. \tag{52}$$

This result is exact for the stated inviscid linear problem, not for the viscous device. With the current parameters, the inviscid curve reaches its growth maximum near $M_s = 0.4398$, with $\omega_2 \simeq 2.207 \times 10^{12}\ \mathrm{s}^{-1}$, and then rolls over. It remains positive as the inviscid sonic endpoint is approached. No continuation of Eq. (52) is assigned to an absent inviscid DC branch.

### 5.3 Weak-growth reduction without a WKB amplitude approximation

An approximate explicit growth increment can also be obtained from the finite-viscosity residual itself. Expand Eq. (46) about a real trial frequency $\omega_r$ close to the target root,

$$\mathrm{R}_\nu(\omega_r + \delta\omega) \simeq \mathrm{R}_\nu(\omega_r) + \delta\omega\ \partial_\omega \mathrm{R}_\nu(\omega_r) = 0. \tag{53}$$

To this order,

$$\omega_1 \simeq \omega_r - \mathrm{Re}\left[\frac{\mathrm{R}_\nu}{\partial_\omega \mathrm{R}_\nu}\right]_{\omega_r}, \qquad \omega_2 \simeq -\mathrm{Im}\left[\frac{\mathrm{R}_\nu}{\partial_\omega \mathrm{R}_\nu}\right]_{\omega_r}. \tag{54}$$

Equation (54) is a Taylor reduction of the full spectral condition. It still requires the nonuniform finite-viscosity propagation, but it does not replace that propagation by local WKB amplitudes. The derivative is taken at fixed DC profile, and the approximation requires a small correction and a nonzero residual derivative. The curves in Fig. 3 use the complete complex root, Eq. (47), rather than this first-order reduction.

### 5.4 Nonuniform two-wave growth expression and its reduction

The earlier two-wave result is useful for identifying the gain and loss terms and is retained here with its approximation stated. Let the real trial frequency $\omega_r$ satisfy the phase part of Eq. (26). Write

$$\begin{gathered} G(\omega_r) = \mathrm{A}(\omega_r) + i\Theta(\omega_r), \qquad \Theta(\omega_r) = 0, \\ \mathrm{A} = \ln\left|\frac{k_+(L)}{k_-(L)}\right| + \mathrm{Im}\int_0^L (k_+ - k_-)\ \mathrm{d}x. \end{gathered} \tag{55}$$

The real quantity A is the logarithmic round-trip amplitude residual. The endpoint term gives the leading DS reflection gain, whereas the propagation integral contains attenuation. Expanding the same complex equation, not a separate empirical gain formula, gives

$$\delta\omega \simeq -\frac{\mathrm{A}}{\partial_\omega \mathrm{A} + i\,\partial_\omega \Theta}. \tag{56}$$

Separating the real and imaginary parts yields the full first-order weak-growth expression within this two-wave construction,

$$\omega_2^{(\mathrm{WKB})} \simeq \frac{\mathrm{A}\,\partial_\omega \Theta}{(\partial_\omega \mathrm{A})^2 + (\partial_\omega \Theta)^2}, \qquad \delta\omega_1 \simeq -\frac{\mathrm{A}\,\partial_\omega \mathrm{A}}{(\partial_\omega \mathrm{A})^2 + (\partial_\omega \Theta)^2}. \tag{57}$$

All quantities in Eq. (57) are evaluated at $\omega_\mathrm{r}$. The commonly used ratio follows only when the frequency derivative of the amplitude residual is small,

$$\omega_2^{(\mathrm{WKB})} \simeq \frac{\mathrm{A}}{\partial_\omega \Theta}, \qquad |\partial_\omega \mathrm{A}| \ll |\partial_\omega \Theta|. \tag{58}$$

The cubic correction can be included at the same approximation level by replacing both $k_+$ and $k_-$ by $\tilde{k}_+$ and $\tilde{k}_-$ in the endpoint ratio, the propagation integral, and their frequency derivatives. Correcting only the phase integral would not give the corresponding corrected growth expression. For a more transparent limit, expand the quadratic roots to first order in the local damping terms. With the branch convention of Eq. (21),

$$\mathrm{Im}\left(k_+^{(Q)} - k_-^{(Q)}\right) \simeq -\frac{\gamma s}{s^2 - v_0^2} - \nu\omega_\mathrm{r}^2 \frac{s^2 + v_0^2}{s(s^2 - v_0^2)^2}. \tag{59}$$

At this order, the endpoint logarithm is $\ln[(s_\mathrm{d} + v_\mathrm{d})/(s_\mathrm{d} - v_\mathrm{d})]$, and $\partial_\omega \Theta \simeq \mathrm{T}$, where T is the dimensional round-trip time in Eq. (33). Substituting these expressions into Eq. (58) gives the nonuniform reduced growth rate,

$$\omega_2^{(Q,\mathrm{red})} \simeq \frac{1}{\mathrm{T}} \ln\left(\frac{s_\mathrm{d} + v_\mathrm{d}}{s_\mathrm{d} - v_\mathrm{d}}\right) - \frac{\gamma}{2} - \frac{\nu\omega_\mathrm{r}^2}{\mathrm{T}} \int_0^L \frac{s^2 + v_0^2}{s(s^2 - v_0^2)^2}\,\mathrm{d}x. \tag{60}$$

The three terms in Eq. (60) are the reflection gain per round-trip time, the momentum-relaxation loss, and the distributed viscous loss. Both the gain time and the loss weighting depend on the nonuniform DC profile. This formula therefore contains more information than adding a drain-local damping rate to a uniform DS gain.

However, Eq. (60) is specifically the reduction of the quadratic two-root dispersion. At finite drift, the omitted cubic term is also linear in viscosity. Consequently, this expression must not be described as the complete first-order viscous expansion of Eq. (18). Appendix C records the first-order change in the bulk attenuation when that cubic term is retained. Neither two-wave reduction replaces the complete nonuniform amplitude propagation in Eq. (47).

### 5.5 Uniform-channel and low-current limits

For a uniform background, Eq. (60) becomes

$$\omega_2^{(Q,\mathrm{uniform})} \simeq \frac{s^2 - v_0^2}{2sL}\ln\left(\frac{s+v_0}{s-v_0}\right) - \frac{\gamma}{2} - \frac{\nu\omega_\mathrm{r}^2(s^2+v_0^2)}{2s^2(s^2-v_0^2)}. \tag{61}$$

The first term is the familiar lossless DS growth rate for a uniform cavity. In the small-current limit, use $\ln[(s+v_0)/(s-v_0)] \simeq 2v_0/s$ and $\omega_\mathrm{r} \simeq q\pi s/(2L)$. Then

$$\omega_2 \simeq \frac{v_\mathrm{s}}{L} - \frac{\gamma}{2} - \frac{q^2\pi^2\nu}{8L^2}, \qquad v_\mathrm{s} \ll s_\mathrm{s}. \tag{62}$$

The source velocity can be used here because the channel is nearly uniform at low current. The cubic-corrected attenuation has the same leading small-current limit. The approximate onset condition is

$$v_{\mathrm{s,on}} \simeq \frac{\gamma L}{2} + \frac{q^2\pi^2\nu}{8L}. \tag{63}$$

Equations (62) and (63) explain why increasing viscosity moves the low-current onset to a larger drift velocity. They are scaling limits, not formulas to be extrapolated to $M_\mathrm{d} \simeq 1$. The restored two-wave expressions in Sections 5.4 and 5.5 are retained to expose the analytical progression. They are not the source of the reported finite-viscosity agreement in Fig. 3.

### 5.6 Comparison with the available growth-rate data

Fig. 3 retains the exact inviscid curve and the full finite-viscosity curves for $\nu = 2$, $20$, and $100\ \mathrm{cm^2/s}$. The finite-viscosity curves are calculated from Eqs. (10), (12), (44), (45), and (47). They are not obtained by inserting the local damping into Eq. (60). All plotted points and curves are carried over from the parameter-updated calculation; no high-current continuation or new contact model is introduced by the present restoration of the derivations.

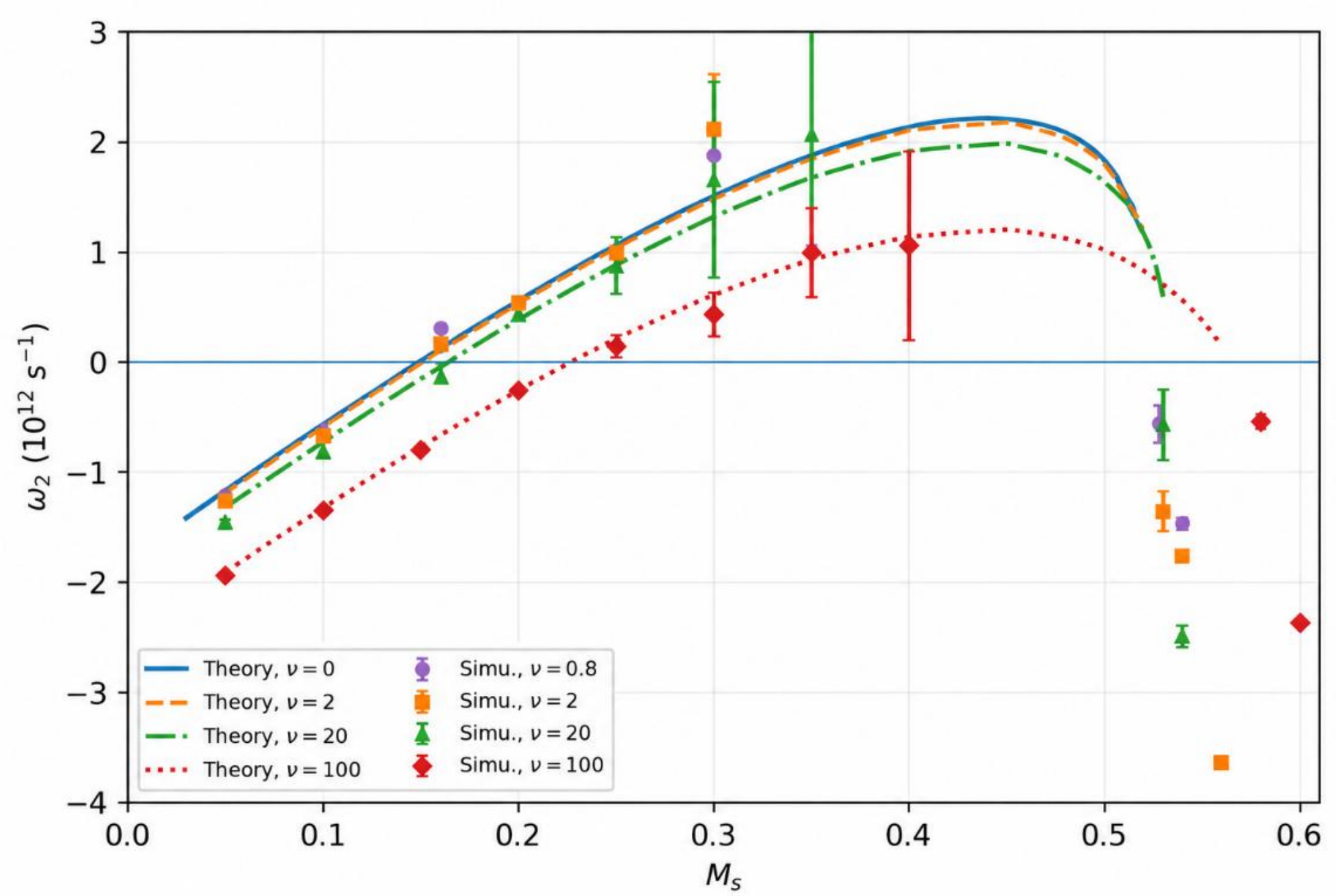

Fig. 3. Growth rate of the fundamental mode. Symbols are the simulation growth-rate data and their reported error bars. The inviscid theory follows Eq. (52). The curves at $\nu = 2$, $20$, and $100$ $\mathrm{cm}^2/\mathrm{s}$ use the complete source-compatible finite-viscosity eigenvalue condition, Eq. (47), on the corresponding viscous DC background. Curves terminate at their previously calculated endpoints; no termination is identified as a mathematical maximum continuation.

Table III retains the quantitative comparison for the available data below $M_\mathrm{s} = 0.5$. The simulated growth rates are sparse: the largest available $M_\mathrm{s}$ is $0.30$, $0.30$, $0.35$, and $0.40$ for the four respective viscosities. The errors therefore support agreement with the available data, not a dense validation of the entire interval up to $0.5$. The approximate reduced formulas above are not assigned these error statistics.

Table III. Reported finite-viscosity eigenvalue comparison with the available subsonic growth-rate data. RMSE units are $10^{12}$ $\mathrm{s}^{-1}$.

| $\nu$ ($\mathrm{cm}^2/\mathrm{s}$) | **Points** | **Largest $M_\mathrm{s}$** | **RMSE** | **Correlation** |
|---|---|---|---|---|
| 0.8 | 6 | 0.30 | 0.179 | 0.990 |
| 2 | 6 | 0.30 | 0.263 | 0.986 |
| 20 | 7 | 0.35 | 0.205 | 0.997 |
| 100 | 8 | 0.40 | 0.079 | 0.998 |

The agreement is strongest for $\nu = 100$ $\mathrm{cm}^2/\mathrm{s}$. For the lower viscosities, a few positive-growth data points contribute appreciably to the error. Some rates were extracted by fitting a limited sequence of transient peaks, and their value can depend on the fitting interval. A statistical fit error does not necessarily include this systematic uncertainty. The comparison consequently supports the finite-viscosity bulk description while not uniquely determining the physical contact condition.

## 6. Finite-viscosity sonic transition and range of validity

The distinction between a regular bulk mode and a contact-selected device mode becomes especially important near the sonic condition. Let $v_* = s_*$ denote the sonic drift velocity on the inviscid trajectory, and define the local slope

$$a_* = v_* \frac{\mathrm{d}}{\mathrm{d}v_0}\left(1 - \frac{s^2(v_0)}{v_0^2}\right)\Bigg|_{\mathrm{v_0=v_*}}. \tag{64}$$

This single coefficient describes the local curvature of the steady characteristic; it is not a rescaling of the channel variables. For the UCCM, direct differentiation of Eq. (11) gives

$$a_* = 3 - \frac{J_\mathrm{d}/\left(C_\mathrm{g}\eta V_\mathrm{T} v_*\right)}{\exp\left[J_\mathrm{d}/\left(C_\mathrm{g}\eta V_\mathrm{T} v_*\right)\right] - 1}. \tag{65}$$

With $\delta v = v_0 - v_*$, the leading terms of Eq. (10) near the sonic position are

$$-\frac{\nu}{v_*}\delta v'' + \frac{a_*}{v_*}\delta v\,\delta v' + \gamma \simeq 0. \tag{66}$$

Balancing these terms gives the velocity and spatial widths

$$\Delta v = \left(\frac{\nu\gamma v_*}{a_*^2}\right)^{1/3}, \qquad \mathrm{L_\nu} = \left(\frac{\nu^2}{a_*\gamma v_*}\right)^{1/3}. \tag{67}$$

The fractional powers show that the sonic transition is a singular perturbation. They do not follow from an ordinary expansion $\omega = \omega^{(0)} + \nu\omega^{(1)} + \cdots$ at a current for which no inviscid steady solution exists. Integrating Eq. (66) once gives a useful dimensional form of the inner equation,

$$\nu\,\delta v' = \frac{a_*}{2}(\delta v)^2 + \gamma v_*(x - x_*). \tag{68}$$

The integration constant has been absorbed in $x_*$. This equation supplies the local Riccati structure without introducing normalized inner coordinates. The numerical steady trajectories remain necessary for the contact-dependent sonic-crossing values.

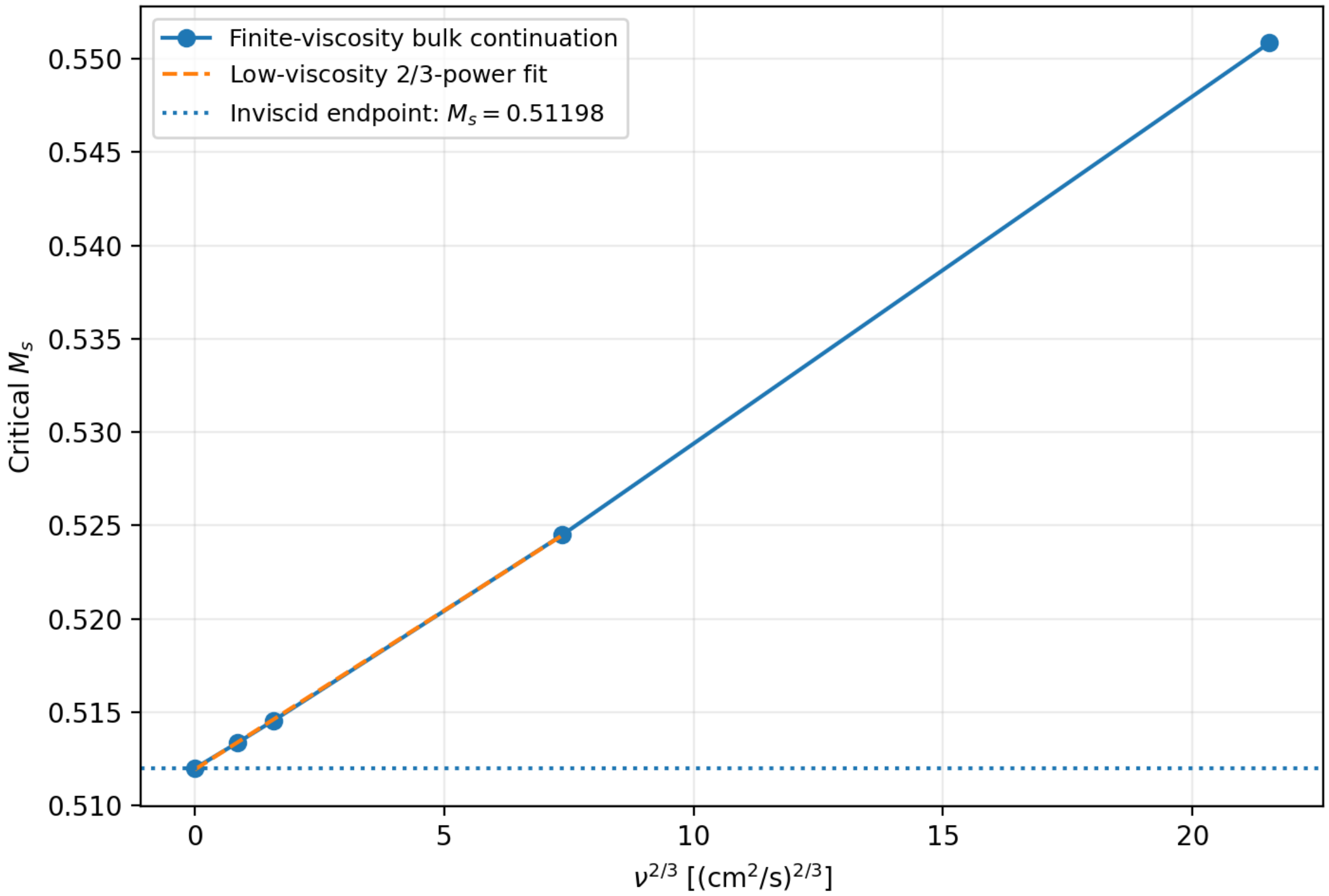

Fig. 4. Source Mach number at the first sonic crossing of the source-compatible bulk branch. The numerical values are $0.51336$, $0.51456$, $0.52449$, and $0.55086$ for $\nu = 0.8$, $2$, $20$, and $100\ \mathrm{cm}^2/\mathrm{s}$. The inviscid value is $0.51198$. The low- and moderate-viscosity shifts approximately follow the $\nu^{2/3}$ dependence associated with Eq. (67). These points are steady-state sonic thresholds, not zeros of $\omega_2$.

The saved velocity boundaries in Eq. (6) require a separate consistency check. For a smooth positive-current DC state, Eq. (10) gives

$$v_0' = 0 \quad \Rightarrow \quad v_0'' = \frac{\gamma v_0}{\nu} > 0. \tag{69}$$

A positive-velocity branch launched from $v_0'(0) = 0$ therefore has positive slope immediately inside the source. Any subsequent zero of the slope could only be crossed from negative to positive. The same smooth branch cannot return to $v_0'(L) = 0$. The finite-viscosity theory in this manuscript consequently uses the source-compatible closure and does not claim to satisfy every active boundary node of the time-dependent simulation.

The available subsonic agreement should not be confused with a proof that the two contact models are identical. Likewise, the negative simulation rates near and above the sonic region cannot yet be assigned uniquely to a sonic layer, contact reflection, or a particular stationary branch. Establishing that correspondence requires the actual stationary spatial fields and the same boundary closure on the DC and AC problems. The present work separates this unresolved comparison from the frequency and growth-rate derivations that are already available.

## 7. Conclusion

In this work, a nonuniform hydrodynamic description was developed for current-driven plasma-wave instability in a gated InGaAs/GaAs channel. The principal results are summarized below.

1) A finite DC drain current drives the electron channel away from a uniform plasma cavity. As carriers accelerate toward the drain, the carrier density and plasma-wave velocity decrease, producing an increasingly nonuniform hydrodynamic state. This spatial redistribution becomes particularly important as the system approaches the sonic regime and strongly modifies the propagation of plasma waves through the channel.

2) The oscillation frequency is therefore governed by the propagation of plasma waves through the entire nonuniform channel rather than by the local conditions at the drain. The current-induced spatial variation changes the effective round-trip propagation time and leads to a pronounced reduction of the oscillation frequency at large current. This behavior is captured naturally by the nonuniform-channel description, whereas a uniform-channel approximation increasingly overestimates the frequency as the sonic regime is approached.

3) The instability can be understood physically as a competition between current-induced amplification and dissipative losses during a plasma-wave round trip. Momentum relaxation and electron viscosity suppress the instability, while the asymmetric source–drain boundary conditions provide the amplification mechanism. Because both wave propagation and dissipation depend on the spatially varying background, the instability threshold and growth rate are intrinsically nonlocal properties of the current-driven channel.

4) Finite electron viscosity has an additional role near the sonic regime: it smooths the singular behavior predicted by the inviscid description and allows the hydrodynamic state to remain regular over a wider range of current. The resulting finite-viscosity theory reproduces the observed subsonic growth rates well. At higher currents, however, the dynamics become increasingly sensitive to the near-sonic structure and to the contact boundary conditions, indicating that the eventual suppression of the instability cannot yet be attributed to bulk hydrodynamics alone.

These results establish current-induced spatial nonuniformity as an essential contributor of plasma-wave instability in realistic gated electron channels, and identify the near-sonic, contact-sensitive regime as the principal remaining challenge for a complete device-level description.

### Appendix A. Quadratic discriminant and first cubic correction

The additional algebra is given here to make the connection between the local dispersion and the frequency hierarchy reproducible. In this appendix only, denote the dimensional square root in Eq. (21) by D. Its real and imaginary squared parts are

$$\mathrm{D}^2 = \omega_\mathrm{r}^2(s^2+\gamma\nu) - \frac{v_0^2\gamma^2}{4} + i\omega_\mathrm{r}(\gamma s^2 - \nu\omega_\mathrm{r}^2). \tag{A1}$$

If $\mathrm{D} = \mathrm{D_R} + i\mathrm{D_I}$, direct division by $s^2 - v_0^2 - i\nu\omega_\mathrm{r}$ gives

$$-\mathrm{Re}\left(k_+^{(Q)} - k_-^{(Q)}\right) = \frac{2[(s^2-v_0^2)\mathrm{D_R} - \nu\omega_\mathrm{r}\mathrm{D_I}]}{(s^2-v_0^2)^2+\nu^2\omega_\mathrm{r}^2}. \tag{A2}$$

Equation (A2) exhibits both the mixed numerator and the finite-$\nu\omega_\mathrm{r}$ denominator of the full quadratic phase. Omitting them produces Eq. (30) only when their contribution and the omitted imaginary-discriminant correction are small.

For the cubic correction, the quadratic-root sum and product are

$$k_+^{(Q)} + k_-^{(Q)} = \frac{v_0(2\omega+i\gamma)}{v_0^2 - s^2 + i\nu\omega}, \qquad k_+^{(Q)}k_-^{(Q)} = \frac{\omega(\omega+i\gamma)}{v_0^2-s^2+i\nu\omega}. \tag{A3}$$

Since the polynomial derivative at the two roots is $\pm 2\mathrm{D}$, their separation changes by

$$\delta k_+ - \delta k_- = \frac{i\nu v_0}{2\mathrm{D}}\left[\left(k_+^{(Q)}\right)^3 + \left(k_-^{(Q)}\right)^3\right]. \tag{A4}$$

The sum of cubes follows from the two identities in Eq. (A3). This is equivalent to expanding the cubic contribution about each root separately. A single Newton step applied to the full cubic would instead contain the extra derivative term $-3i\nu v_0\left(k_\pm^{(Q)}\right)^2$ in the denominator; that partially resummed correction should not be confused with the strict first correction used in Eq. (23).

### Appendix B. Closed frequency reductions and the finite-friction term

For the strong-inversion inviscid profile, the Level II integral becomes

$$\int_0^L \frac{\sqrt{s^2+\gamma\nu}}{s^2-v_0^2}\,\mathrm{d}x = \frac{2}{3\gamma}\int_{\mathrm{M_s}}^{M_\mathrm{d}} \frac{\sqrt{1+(\gamma\nu/v_\mathrm{cr}^2)M^{2/3}}}{M^2}\,\mathrm{d}M. \tag{B1}$$

The required antiderivative is verified by direct differentiation,

$$\int \frac{\sqrt{1+(\gamma\nu/v_{\mathrm{cr}}^2)M^{2/3}}}{M^2}\,\mathrm{d}M = -\frac{[1+(\gamma\nu/v_{\mathrm{cr}}^2)M^{2/3}]^{3/2}}{M}. \tag{B2}$$

Taking the endpoint difference gives Eq. (37). No additional dimensionless substitution is needed. To retain the finite-$\gamma^2$ correction before expanding the resulting frequency, define two dimensional integrals only for this step,

$$I_{\mathrm{ph}} = \int_0^L \frac{\sqrt{s^2+\gamma\nu}}{s^2-v_0^2}\,\mathrm{d}x, \qquad I_{\mathrm{fr}} = \int_0^L \frac{v_0^2}{(s^2-v_0^2)\sqrt{s^2+\gamma\nu}}\,\mathrm{d}x. \tag{B3}$$

Both integrals have units of time. The square-root expansion in Eq. (30) gives

$$2I_{\mathrm{ph}}\omega_1 - \frac{\gamma^2 I_{\mathrm{fr}}}{4\omega_1} \simeq q\pi. \tag{B4}$$

The positive root is

$$\omega_1 \simeq \frac{q\pi + \sqrt{q^2\pi^2 + 2\gamma^2 I_{\mathrm{ph}} I_{\mathrm{fr}}}}{4I_{\mathrm{ph}}}. \tag{B5}$$

Expanding Eq. (B5) once more gives Eq. (38). At $\nu = 0$ on the strong-inversion profile, $I_{\mathrm{fr}} = 2(M_{\mathrm{d}} - M_{\mathrm{s}})/(3\gamma)$, giving Eq. (39). Thus the numerical Level III integral, its positive-root approximation, and its first explicit correction are separate levels of reduction, rather than interchangeable formulas.

**Appendix C. Attenuation, cubic consistency, and the weak-growth derivative**

The attenuation formula in Eq. (59) follows by expanding the square root and quadratic denominator in Eq. (21) to first order in $\gamma$ and $\nu$. It is the result retained in the earlier quadratic gain–loss derivation. A useful consistency check is obtained by performing the same expansion directly on the full local dispersion, Eq. (18). With $k_{\pm}^{(0)} = -\omega_{\mathrm{r}}/(s-v_0)$ and $\omega_{\mathrm{r}}/(s+v_0)$ for the upstream and downstream waves, respectively, the first local damping corrections are

$$\mathrm{Im}k_+ \simeq -\frac{\gamma + \nu\omega_{\mathrm{r}}^2/(s-v_0)^2}{2(s-v_0)}, \qquad \mathrm{Im}k_- \simeq \frac{\gamma + \nu\omega_{\mathrm{r}}^2/(s+v_0)^2}{2(s+v_0)}. \tag{C1}$$

Subtraction gives

$$\mathrm{Im}(k_+ - k_-) \simeq -\frac{\gamma s}{s^2-v_0^2} - \nu\omega_{\mathrm{r}}^2\frac{s(s^2+3v_0^2)}{(s^2-v_0^2)^3}. \tag{C2}$$

Equations (C1) and (C2) are an algebraic expansion of the already retained cubic dispersion, not a new numerical fit. The difference from Eq. (59) is of order $\nu v_0^2$ away from the sonic point. It explains why a two-root quadratic attenuation formula should not be labeled the complete first-order finite-viscosity result at appreciable drift.

Keeping the same leading endpoint gain and phase time, this full-dispersion bulk attenuation would give the diagnostic reduction

$$\omega_2^{(\mathrm{cubic,red})} \simeq \frac{1}{\mathrm{T}} \ln\left(\frac{s_\mathrm{d}+v_\mathrm{d}}{s_\mathrm{d}-v_\mathrm{d}}\right) - \frac{\gamma}{2} - \frac{\nu \omega_\mathrm{r}^2}{\mathrm{T}} \int_0^L \frac{s(s^2+3v_0^2)}{(s^2-v_0^2)^3}\,\mathrm{d}x. \tag{C3}$$

Equation (C3) still neglects exact nonuniform amplitude transport and any contact contribution of the third root. It is not asserted to reproduce Fig. 3, and it is not extrapolated to $s = v_0$. Both Eqs. (60) and (C3) recover the small-current limit in Eq. (62).

For completeness, the derivative of a full boundary residual can be evaluated without a finite-difference shift of the frequency. If the three physical perturbation variables are assembled into $\boldsymbol{\psi} = (n_1, v_1, v_1')^\mathrm{T}$ and Eqs. (44)–(45) are denoted by $\boldsymbol{\psi}' = \mathbf{A}(x,\omega)\boldsymbol{\psi}$, differentiation gives

$$(\partial_\omega \boldsymbol{\psi})' = \mathbf{A}\,\partial_\omega \boldsymbol{\psi} + (\partial_\omega \mathbf{A})\boldsymbol{\psi}. \tag{C4}$$

The derivative of the source-normalized initial data is zero when $v_\mathrm{ref}$ is frequency independent. At the drain,

$$\partial_\omega \mathrm{R}_\mathrm{v} = e[v_\mathrm{d}\,\partial_\omega n_1(L) + n_\mathrm{d}\,\partial_\omega v_1(L)]. \tag{C5}$$

These equations are a direct differentiated form of the stated linear boundary problem. They explain how Eq. (54) can be evaluated consistently at fixed background. They do not introduce a new physical fitting parameter or replace the full-root results used in the figures.

**ACKNOWLEDGMENTS**

The author gratefully acknowledges the late Professor Michael S. Shur for his invaluable guidance and insightful discussions that contributed to the theoretical analysis presented in this work, and for his mentorship and support throughout the author's doctoral studies. This work is dedicated to his memory.

**AUTHOR DECLARATIONS**

The author has no conflict-of-interest to disclose.

**DATA AVAILABILITY**

The simulation and analytical data are available from the author upon reasonable request.

## REFERENCES

[1] M. Dyakonov and M. Shur, “Shallow water analogy for a ballistic field effect transistor: New mechanism of plasma wave generation by dc current,” Phys. Rev. Lett. 71, 2465–2468 (1993).

[2] A. P. Dmitriev et al., “Numerical study of the current instability in a two-dimensional electron fluid,” Phys. Rev. B 55, 10319 (1997).

[3] M. V. Cheremisin, M. I. Dyakonov, M. S. Shur, and G. Samsonidze, “Influence of electron scattering on current instability in field effect transistors,” Solid-State Electronics 42, 1737–1742 (1998).

[4] M. V. Cheremisin and G. G. Samsonidze, “D’yakonov–Shur instability in a ballistic field-effect transistor with a spatially non-uniform channel,” Semiconductors 33, 619–628 (1999).

[5] C. B. Mendl and A. Lucas, “Dyakonov–Shur instability across the ballistic-to-hydrodynamic crossover,” Appl. Phys. Lett. 112, 124101 (2018).

[6] C. B. Mendl, M. Polini, and A. Lucas, “Coherent terahertz radiation from a nonlinear oscillator of viscous electrons,” Appl. Phys. Lett. 118, 013105 (2021).

[7] Y. Zhang and M. S. Shur, “THz detection and amplification using plasmonic field effect transistors driven by DC drain currents,” J. Appl. Phys. 132, 193102 (2022).

[8] Y. Zhang and M. S. Shur, “TeraFET terahertz detectors with spatially non-uniform gate capacitances,” Appl. Phys. Lett. 119, 161104 (2021).

[9] Y. H. Byun, K. Lee, and M. Shur, “Unified charge control model and subthreshold current in heterostructure field-effect transistors,” IEEE Electron Device Lett. 11, 50–53 (1990).

[10] COMSOL Multiphysics® Ver. 5.4, COMSOL AB, Stockholm, Sweden.

[11] M. I. Dyakonov and M. S. Shur, “Choking of electron flow: A mechanism of current saturation in field-effect transistors,” Phys. Rev. B 51, 14341–14345 (1995).

[12] M. Dyakonov and M. Shur, “Detection, mixing, and frequency multiplication of terahertz radiation by two-dimensional electronic fluid,” IEEE Trans. Electron Devices 43, 380–387 (1996).

[13] M. I. Dyakonov and M. S. Shur, “Plasma wave electronics: Novel terahertz devices using two-dimensional electron fluid,” IEEE Trans. Electron Devices 43, 1640–1645 (1996).

[14] W. Knap et al., “Terahertz emission by plasma waves in 60 nm gate high electron mobility transistors,” Appl. Phys. Lett. 84, 2331–2333 (2004).

[15] W. Knap et al., “Field effect transistors for terahertz detection: Physics and first imaging applications,” J. Infrared Millim. Terahertz Waves 30, 1319–1337 (2009).

[16] I. Torre, A. Tomadin, A. K. Geim, and M. Polini, “Nonlocal transport and the hydrodynamic shear viscosity in graphene,” Phys. Rev. B 92, 165433 (2015).

[17] D. A. Bandurin et al., “Negative local resistance caused by viscous electron backflow in graphene,” Science 351, 1055–1058 (2016).

[18] J. Crossno et al., “Observation of the Dirac fluid and the breakdown of the Wiedemann–Franz law in graphene,” Science 351, 1058–1061 (2016).

[19] P. J. W. Moll et al., “Evidence for hydrodynamic electron flow in PdCoO2,” Science 351, 1061–1064 (2016).

[20] A. Principi, G. Vignale, M. Carrega, and M. Polini, “Bulk and shear viscosities of the two-dimensional electron liquid in a doped graphene sheet,” Phys. Rev. B 93, 125410 (2016).

[21] A. Lucas and K. C. Fong, “Hydrodynamics of electrons in graphene,” J. Phys.: Condens. Matter 30, 053001 (2018).

[22] M. Polini and A. K. Geim, “Viscous electron fluids,” Phys. Today 73(6), 28–34 (2020).